# Mediation with near-insolvent defaulting suppliers :
# a linear optimisation model to find an optimal outcome

Eric Lavallée

*Lavery, de Billy llp*

95-200 Boul. Jacques-Cartier Sud, Sherbrooke, Quebec, Canada, J1J 2Z3, elavallee@lavery.ca

**ABSTRACT**

This paper presents a model to describe contractual dispute resolution by mediation in situations where a defaulting supplier is near-insolvent. While each party has internal constraints, and if alternate performances are available, such as more costly alternative goods, the proposed approach allows the mediator to find an optimal solution. The notion of optimality is presented as adherence to the initial contract, therefore optimising a value function for the non-defaulting party. The proposed model includes describing the evolution over time of each party's perceived constraints using a phasor-like approach with a modulation to the core constraints phasing out of the real part and phasing in the imaginary part of complex numbers. The offers related to alternative performances by the defaulting party are modelled by a Gompertz function, being an exponential learning curve of the supplier in regards to the reaction to its offers, limited by another exponential function when approaching its internal constraints. Furthermore, the model takes into account the discount associated to the delay in the delivery time of the alternative performances.

# 1. INTRODUCTION

Contractual dispute resolution is often required when a party is in default to its contract or when there is a disagreement on the execution of the obligations of each parties. Parisi et al. (2011) have demonstrated, for bilateral contracts, that it is possible to correct imperfections in contract enforcement, such as imperfect compensation due to parties' insolvency. While recourse to courts or arbitration is common in such cases, the parties may be willing to try to resolve the dispute by mediation to settle the issue quickly and avoid legal expenses. Mediation to resolve contractual disputes is a form of negotiation, and therefore a non-cooperative game.

Wilson (2007), showed that, with a mediator making sequential random proposals, the subgame perfect payoff converges to an asymmetric Nash solution with weights influenced by the relative discount rates of the players. Others have suggested adding noise to the information gathered by the mediator (Goltsman et al. 2009). However, real-life mediators usually have a strategy and the proposals are not random.

Ross and Chen (2007) have studied experimentally the effect of online mediator strategy, but their results show no effect of the mediator strategy over the concessions the parties offer. In their case however, the parties had pre-programmed to make moderately competitive proposals. On the other hand, Ethano et al. (2001) have proposed a step-by-step method to reach a Pareto-efficient agreement. Their approach does not require the parties to reveal their utility functions (constraints). In their approach, the mediator helps the parties find favourable directions to the negotiation.

Zeleznikow and Bellucci (2006) have made proposals for a negotiation decision support system to support mediators to dynamically modify initial preferences throughout the negotiation process. However, in commercial contractual dispute resolution, it might not be the preferences that shift during the mediation process, but rather the perception of their own constraints by each party. The mediator's role includes acting as an "agent of reality" and to "increases the parties' perceptions of their cases" (AAA, 2013). Each party may reassess its constraints during the mediation process, a form of introspection, leading usually to more reconcilable positions for each party over time.

Therefore, it is important to propose an approach allowing the mediator to identify an optimal solution which takes into account the constraints of the parties, the evolution of the perception by each party of their constraints, and an approach allowing a mediator to guide the parties to a settlement which is as close as possible to an optimal solution.

Also, it is needed to have a model taking into account the evolution of the offers that will be made by the defaulting supplier during the mediation process. In a defaulting supplier mediation, the parties are however limited by their internal constraints, and the relation is therefore not simply exponential nor linear. Offers of the defaulting party should tend asymptotically to a maximum situation that would affect its economic viability. Montrol (1978) has suggested that almost all social phenomena can be approximated by such a sigmoid function, except for unusual intermittent events. Foster and Wild (1999) have demonstrated the use of logistic functions for econometric modelling, and Li et al. (2006) have suggested it as a model for static concessions in automated e-business negotiations. A sigmoid function may be inadequate to describe Boulwarism, whereas a party will make a "take-it or leave-it" offer, which is out of the scope of this paper. Among the sigmoid functions, the logistic functions represent a broad family of functions. Sánchez-Chóliz and Jarne (2015) have studied the limitations of the logistic and Gompertz functions, which both have functional weaknesses, noting however, their operational advantages. The present paper proposes using the Gompertz function to describe the evolution of the offers in a mediation.

This paper also describes a method whereas an optimisation technique may be used by a mediator to determine an optimal outcome, and also provide a model to describe a typical mediation process. In order to provide such description, we must consider the evolution of each party's perception of their own constraints, as well as the discount rate associated with delays in delivery.

## 2. PRISONER'S DILEMMA AND OPTIMISATION

In a contract where a supplier fails to delivers the amount and/or type of goods expected by the buyer, and where the buyer retains payment of the sums agreed upon in retaliation for the undelivered goods, each party may make compromises in order to resolve the issuing dispute. The supplier may want to provide alternative performance in order to get paid. Conversely, the buyer may want to pay the amounts due in order for the deliveries to resume. Depending on the contract, the buyer may have other means of putting pressure on the defaulting supplier, such as threatening to exercise a penalty clause or preventing the other party from getting other contracts.

During conventional bipartite negotiation, none of the parties may be willing to make concession and take the first step towards resolution, for fear of obtaining nothing or little in exchange. Therefore, no party is willing to exchange their real position, leading to a standstill of the negotiation process, and possibly to a long and costly litigation process.

The utility of the negotiation process is therefore limited by the fact that each party ignores the other party's real position. It is therefore a prisoner's dilemma, since both party would be better of by divulging their real position to the other party, but is unwilling to make the first move toward a solution.

Furthermore, mediation usually occurs in situations whereas each party have constraints, but is not free to "shop around" to find a new partner. Alternative supplier may be considered, but may come at a cost due to delays, excess in prices, etc. Contrarily to a normal negotiation process, the mediation process is therefore bounded by constraints.

In modern mediation process, the mediator usually attempts to understand the real position of each party, conducting interviews with each of them individually. Such talks are usually confidential, and the mediator will only divulge to the other party what he is authorised to.

However, getting both positions and both sides of the story, the mediator is able to find a solution whereas both parties make compromises (Klein et al. 2002).

The following table illustrates the various possible outcomes of the negotiation process described above as a prisoner's dilemma :

|  | Compromise by A | No compromise by A |
|---|---|---|
| Compromise by B | *optimal solution, leads to a settlement agreement* | *sub-optimal for B, possibly non-acceptable* |
| No compromise by B | *sub-optimal for A, possibly non-acceptable* | *no solution, leads to litigation* |

Several jurisdiction around the world have adopted policies implementing alternative dispute resolution methods, including mediation. In commercial contracts between a buyer and supplier, mediation can be especially useful to resolve situations where the buyer needs to receive a certain amount of goods, but the supplier is unable to supply the full contracted amount of goods. However, contractual stability needs to be preserved as much as possible. Indeed, several legal systems around the world have rejected the "unforeseeability" theory. For example, the *Code civil* (France) provides that: "Contracts legally formed are the law of the parties that made it. They cannot be revoked unless from the mutual consent, or by causes authorised by law. They must be performed in good faith."[1] The *Civil Code of Quebec* provides that: "A contract may not be resolved, resiliated, modified or revoked except on grounds recognized by law or by agreement of the parties"[2]. Common law jurisdiction have historically rejected unforeseen changes as a justification of default, except in cases of frustrated contracts[3]. In general, contracts cannot be modified because of hardship of a single party, unless with the consent of the other parties.

On the other hand, the UNIDROIT principles include both an article on the binding character of the contract as a general rule[4], and provisions allowing the disadvantaged party to request renegotiations in limited cases of hardship[5]. Similar notions are integrated in the German civil code (*Bürgerliches Gesetzbuch*), whereas a party may demand adaptation of the contract, or even revoke the contract if circumstances which became the basis of a contract have significantly changed[6]. The purpose of such adaptation is to restore equilibrium between the parties. Furthermore, the notion of "good faith" is used in several legal systems to entice a non-defaulting party to collaborate in finding solutions to the other party's unforeseen constraints, or at least not to exercise contractual rights abusively against such other party. For example, the *Uniform Commercial Code,* in force in almost every U.S. states, provides that: "Every contract or duty within the Uniform Commercial Code imposes an obligation of good faith in its performance and enforcement."[7]

---

[1] s. 1134; translation of : "*Les conventions légalement formées tiennent lieu de loi à ceux qui les ont faites. Elles ne peuvent être révoquées que de leur consentement mutuel, ou pour les causes que la loi autorise. Elles doivent être exécutées de bonne foi.*"
[2] s. 1439.
[3] *Taylor & Anor v Caldwell & Anor,* [1863] EWHC QB J1 (6 May 1863).
[4] Art. 6.2.1.
[5] Art. 6.2.2 and 6.2.3.
[6] s. 313.

The parties to an agreement usually enter such agreement expecting the other parties and themselves to abide to the contract, this being an essential consideration to the contract itself. In a defaulting supplier situation, mediation may be used to counter the occurrence of unforeseen constraints. The goal of mediation should therefore be to reach not only an efficient solution, but also a solution which promotes contractual stability, therefore adherence to the contractual obligations of the parties. The mediator does not only need to find a mutually acceptable solution to both parties, but should also, whenever possible, favour solutions which have this added notion of "optimality" issuing from the compliance to the initial contract requirements.

In a broader perspective, free contracts could not exist if there was no contractual stability. The optimal solution has a social utility beyond resolving the dispute between two parties, which includes the social reliance on contractual systems and the avoidance of courts overcrowding to resolve disputes that can be solved otherwise.

The mediator should therefore be biased toward contractual stability, which can be assimilated to a bias toward the non-defaulting party. Ivanov (2009) has demonstrated that such bias in a strategic mediator provides the high payoff solution, as if the parties had communicated through an optimal non-strategic mediator. In the proposed model, this bias toward contractual stability is introduced by linear optimisation of a function which is defined using contractual values to the non-defaulting party.

## 3. THE MODEL

3.1 Definitions and proposition

Let's consider a situation whereas the supplier is unable to fulfil its obligations towards a buyer regarding the delivery of a number of units.

Let's first define a value function of each type of alternative performance the defaulting supplier may provide to compensate for the obligation it is unable to fulfil. Under this function, there are $p$ alternative performances types possible by the defaulting supplier. There are $q$ possible moments delivery of said alternative performances (instalments):

$$F(z_{n,m}) = \sum_{n=1}^{p} \sum_{m=1}^{q} V_n\, D(u_m)\, z_{n,m} \qquad (1)$$

whereas $\forall\, n\ \exists\ C_n$

---
[7] § 1-304; see also § 2-615, which provide excuse by failure of presupposed conditions in sales.

$V_n$ being the contractual value of the performance for the buyer, and $C_n$ the corresponding cost for the defaulting supplier. For monetary performances, such as payment reductions or reimbursements, $V_n \in [0,1[$ since there is an internal cost to find and alternate supplier and/or to manage the reimbursement. $u_m$ is the delay of delivery corresponding to a moment *m*. $D(u_m)$ is the discount ratio function attributable to the delay of delivery (see section 3.2 below). We suggest using complex numbers for $z_{n,m}$ to take into account the evolution by each party of its perceived constraints (see section 3.3 below)

A set of constraints for each party shall be enunciated using these $z_{n,m}$, $V_n$ and $C_n$.

For near-insolvent defaulting suppliers, a solution from linear optimisation of $F(z_{n,m})$ with this set of constraint will correspond to an outcome which is (i) acceptable to both parties, and (ii) promote contractual stability.

Indeed, if we define the near-insolvent situation to be a situation where **max** $F(z_{n,m})$ < value of the obligations contracted by the supplier in the initial contract (i.e. that the supplier is not able to provide alternative performance having substantially the same value as the initial contract.) Element (i) of proposition 1 is inferred from being a solution within the range of the constraints for both parties. Element (ii) results from that the optimisation provides the highest $F(z_{n,m})$ value, therefore the closest total value of performances possible for the non-defaulting party.

Let's take the situation of a supplier having to deliver a certain amount of similar units (products such as computers, other devices or furniture). In cases where the supplier is unable to deliver the contracted number of units as prescribed by the initial contract, there is a number of information the mediator may gather by meeting the parties individually, under confidentiality:

*From the supplier*

- cost of contractual units for the supplier ($C_{init}$);
- cost of an alternative type of units the supplier could procure for each instalment ($C_{alt, m}$);
- maximum number of units of alternative type available on the market, for each possible instalment ($N_{max, m}$);
- maximum aggregate value of alternative units the supplier is able to buy due to its solvency limit during the performance of the contract (S);
- maximum aggregate cost of alternate performance (including price reductions) the supplier is able to accept due to its solvency post-performance of the contract (R);

*From the buyer*

- number of undelivered units required by the buyer to operate ($N_{min}$); and
- value of the various performances offered by the supplier ($V_n$).

## 3.2 Discount for delivery delays

The delivery delay for the alternative performance usually causes the buyer to discount them, and therefore it is necessary to determine a discount function for the various instalments. For the purpose of this model, as simple hyperbolic discount function is a good approximation of the perception of the buyer. A hyperbolic discount function has been described by others (Mazur 1987; Kirby 1997) as :

$$D(u_m) = \frac{1}{1 + ru_m} \quad (2)$$

where $r$ is the discount rate and $u_m$ is the delay for the delivery of the alternative performance by the defaulting supplier. The rate $r$ expresses the confidence level of the buyer toward the supplier; if the buyer has a low level of confidence, the rate $r$ will be high, and the value of delayed performances will be much lower than immediate performance. We can imagine cases whereas the rate $r$ will be high enough that delayed performances will have little or no value to the buyer, and that optimisation will rely almost solely on immediate performance of the defaulting supplier.

Mazur (1987) mentions an hyperbolic functions with exponents as sometimes more adequately describing of discount behaviour. While a simple hyperbolic function is used in the present model for simplicity, the model would also be compatible with a more complex functions.

## 3.3 Perceived constraints and time dependency

In post-contractual negotiations, parties may find themselves progressively renouncing to some of their contractual rights to focus on essential requirements. The constraints evolve over time. Newell et al. (2001) showed the use of complex numbers for time scale of changes in the case of motor learning, and Saaty (2007) for time-dependant decision making.

Likewise, in this model, the perceived constraint is the real part of a complex function being a phasor-like modulation as a function of time, added to a fixed core value. Once the core value is reached, the constraint remains static (fully assessed). Under the proposed phasor-like method, the modulation will phase-out of the real part over time during the mediation process. Hereunder, core constraints values are marked "core" and modulation are marked "mod".

$$N_{min}(t) = N_{core} + (N_{mod}\exp(ikt)) \quad \text{for } 0 < kt \leq \pi/2 \text{ and} \quad (3)$$
$$N_{min}(t) = N_{core} + iN_{mod} \quad \text{for } kt > \pi/2$$

where typically $N_{core} + N_{mod} \cong$ initially contracted number of units; and

where $t$ is the elapsed mediation time, on a $\pi/2$ scale; and also

$$R(t) = R_{core} + (R_{mod}\exp(iqt)) \quad \text{for } 0 < qt \leq \pi/2 \text{ and} \quad (4)$$
$$R(t) = R_{core} + iR_{mod} \quad \text{for } qt > \pi/2$$

$$S(t) = S_{core}+(S_{mod}\exp(iqt)) \quad \text{for } 0 < qt \leq \pi/2 \text{ and} \quad (5)$$
$$S(t) = S_{core}+iS_{mod} \quad \text{for } qt > \pi/2$$

This method could be generalised to describe more complex situations. For example, there could be situations where a party could be enticed to go beyond its core constraints, corresponding to variations for time values above π/2. This could give, between $t=0$ and $t=\pi$, a sigmoid shape to the position of the party, as often seen in social phenomenons. A party might also initially under-estimate its core constraint, leading to a phase offset for the initial situation. There could also be, in different cases, a plus and minus uncertainty around the core value, for example if a party is unable to determine accurately its core constraints; $kt$ and/or $qt$ can be unbounded and the value $\text{Re}\{z_x\}$ will oscillate around the core value. Such oscillation would lead to uncertainty in the resulting optimal settlement solution.

For the purpose of this paper, we shall only consider situations of fully assessable core constraints: the core constraints corresponds to the real minimal or maximal values associated to these constraints, and there will not be any negative $\text{Re}\{\exp(it)\}$ reversing the modulation of the constraints, as per the above set of constraints time dependency functions (3), (4) and (5).

It is to be noted that the modulation phase-out ratio $k$ and $q$ may be different for the buyer and the supplier, depending on the negotiating representatives personality and strategy.

3.4 Evolution of offers over time

In general, the defaulting supplier will likely enter the mediation process by expressing a position without concession on terms that are not foreseen of the initial contract (the "initial stage"). To avoid litigation, it will then make rapid concessions (the "intermediary stage"), and end with final stabilisation whereas it will be unable to concede beyond its internal constraints (the "final stage"). This progression can be approximated by a sigmoid shape, in Figure 1.

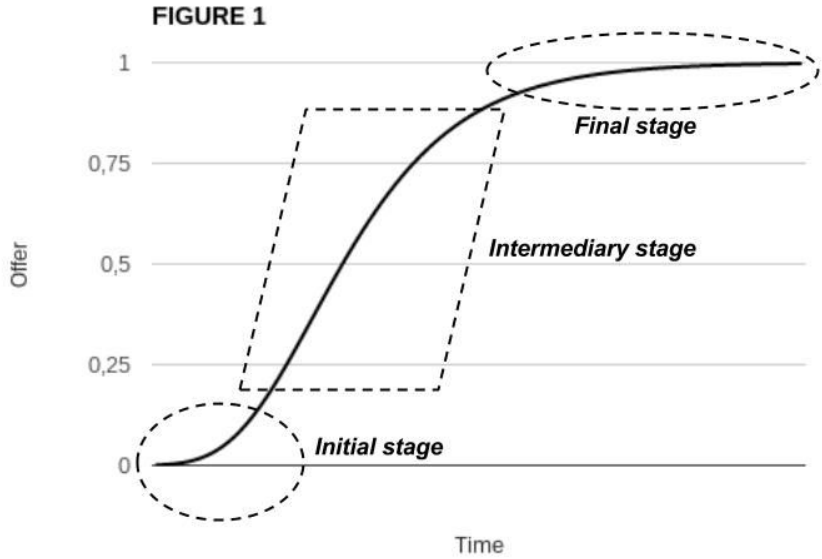

FIGURE 1

The alternative units and the reductions in payment are not terms of the initial contract. For such elements, there will be an initial stage whereas the defaulting party might not consider making offers of such alternative performance. The defaulting supplier is expected to present offers progressively in an intermediary stage, until being limited by internal constraints in the final stage. Sigmoid functions are known to describe learning by trials and errors (Leibowitz et al. 2010), and more specifically negotiation positions (Li et al. 2006; Gal & Pfeffer 2006).

In this model, offers can be modelled as an exponential learning curve limited by an embedded exponential corresponding to the constraints, hence a Gompertz function:

$$g_{n,m} = \text{Re}\{z_{n,m}\} \exp(-a\, e^{-bt}) \qquad (6)$$

Parameter $a$ corresponds to the delay to start making concessions and parameter $b$, the rate of such concessions. Therefore, the value of the offers received, from the buyer standpoint, is:

$$G(g_{n,m}) = \sum_{n=1}^{p} \sum_{m=1}^{q} V_n\, D(u_m)\, g_{n,m}$$

And therefore:

$$G(z_{n,m}) = \text{Re}\{\sum_{n=1}^{p} \sum_{m=1}^{q} V_n\, D(u_m)\, z_{n,m}\} \exp(-a\, e^{-bt}) \qquad (7)$$

where $t$ is the elapsed mediation time

This function $G(z_{n,m})$ can both serve the function of determining the value of the offered alternative performance with the initial contractual value. It can also be used to compare the offer of the defaulting supplier with alternate supplier, if they are available. This allows the buyer to determine if the value of the proposed outcome with the supplier makes this proposal acceptable in regards to alternative solutions, such as seeking alternate suppliers to provide the required goods or services.

For example, lets define an alternate supplier's offer, using the same notation, as the following function :

$$H(z_{l,k}) = \sum_{k=1}^{p} \sum_{l=1}^{q} V_l \, D(u_k) \, z_{l,k} \qquad (8)$$

The discount ratio $r$ for the alternative supplier may be different than for the defaulting supplier. For example, if a long and usually good relationship exist between the buyer and the defaulting supplier, this ratio $r$ may be lower for the defaulting supplier and the buyer will be willing to wait for delayed performance in order to keep its usual supplier. A low level of trust toward the alternate supplier would yield to the same result. At the other extreme, the ratio $r$ may be higher for the defaulting supplier in cases where there is no relation of trust between the buyer and the defaulting supplier, or if the nature of the default is such that it has annihilated the confidence of the buyer in the defaulting supplier.

If there is such alternate supplier, the buyer should accept the settlement only if the following condition is met:

$$G(z_{n,m}) > v \, H(z_{n,m}) + w \, A \, D(u_A) - B \qquad (9)$$

where A is the amount of the damages that the supplier is expecting to be awarded in the event a lawsuit is instituted by the buyer against the defaulting supplier (taking into account what the supplier believes to be solvency limits of the defaulting supplier), $u_A$ being the amount of time expected for the damages to be recuperated from the defaulting supplier, $w$ a risk factor $\in [0,1]$ corresponding to the risk of losing in court in the event of litigation with the defaulting supplier, $v$ a risk factor $\in [0,1]$ corresponding to the risk of the new supplier being also unable to fulfil the contractual requirements, and B is an anticipated cost of terminating the contract of with the defaulting supplier. The later is a compound of many factor, depending on the terms of the contract and other circumstances. It may include, for example, legal and management fees, contractual penalties to third parties and internal costs associated to doing business with a new supplier. A study of the evaluation of third party alternative could provide more details on these variables, but is outside of the scope of this paper.

3.5 Simple case

Let's take the case of alternative performances where the supplier is able to deliver a set of alternative units $z_{1,m}$ and a reimbursement in a single instalment $z_{2,1}$. Making the assumptions that :

- the value of the alternative units $z_{1,1}$ and $z_{1,2}$ is substantially the same value $V_1$;
- these alternative units are replacement performances for units that had an initial value of $V_{init}$;
- the value of reimbursement is depreciated, $V_2 \in [0,1[$ since there is an internal cost of managing the reimbursement and finding an alternate supplier;
- there are two instalments of alternative units ($q=2$);
- the cost per alternative unit will remain constant throughout the duration of all the instalments, $(C_{alt,1} = C_{alt,2} = C_{alt})$ ;
- there is a single instalment of the reimbursement ($m=1$); and
- there is a fixed maximum of units $N_{max,1}$ for the first instalment.

Optimal solution may be found by the linear optimisation of :

$$F(z_{n,m}) = ( \sum_{m=1}^{2} V_1 D(u_m) z_{1,m} ) + (V_2 D(u_1) z_{2,1} ) \qquad (10)$$

considering:

$$Re\{ \sum_{m=1}^{2} z_{1,m}\} \geq Re\{N(t)\}$$

$$Re\{ \sum_{m=1}^{2} z_{1,m}\} \leq N_{max,m}$$

$$Re\{ (C_{alt}-C_{init}) \sum_{m=1}^{2} z_{1,m})+ z_{2,1} \} \leq Re\{R(t)\}$$

$$Re\{ \sum_{m=1}^{2} C_{alt} z_{1,m} \} \leq Re\{S(t)\}$$

$$Re\{z_{n,m}\} \geq 0 \quad \forall \; n, m$$

The above constraints above are inequalities on the real part of the functions, since only this part is relevant to the determination of the optimal solution at a given time.

Without a settlement through mediation, the supplier may go bankrupt, which would leave the buyer needs unsatisfied. The supplier may deliver a number of alternative performance to the contract. Typically, if there exist a less costly alternative ($C_{alt} < C_{init}$), there will be no default, or at least no dispute since the supplier will not increase its solvency problem but rather decrease it by delivering alternative units. However, when, if the alternatives are more costly than the initial contracted units ($C_{alt} > C_{init}$), the buyer may be reluctant to supply the required amount of units as he will lose money on each alternative unit delivered, at least compared to his contractual expectations.

The best possible outcome for the buyer is is to obtain as much undelivered units as possible $z_{1,m}$, as it would be more costly for the buyer to procure independently those units then the initial contractual value per unit, and to compensate undelivered units by reimbursements $z_{2,m}$.

For the purpose of this simple case, near-insolvency situation could be defined as:

$$\text{Re}\{S(t)\} < \text{Re}\{C_{init}N_1 + C_{alt}(N_{min}(t) - N_{max,1})\} \tag{11}$$

This corresponds to situations where $N_{max,2}$ is limited by the solvency parameter $S(t)$ of the supplier. Also, for the purpose of this simple case, lets define the following boundary, meaning that the discount due to the delay in the delivery of the second instalment of alternative units does not causes a decrease in its value to the buyer beyond the point where reimbursement is preferable:

$$\frac{V_1/V_{init}}{1+ru_2} > \frac{V_2}{1+ru_1} \tag{12}$$

The optimisation of $F(z_{n,m})$ gives the following result:

$$z_{1,1} = N_{max,1} \tag{13}$$

$$z_{1,2} = (S(t) - C_{alt}N_{max,1})/C_{alt}$$

$$z_{2,1} = R(t) - [(C_{alt} - C_{init}) \times (N_{max,1} + (S(t) - C_{alt}N_{max,1})/C_{alt})]$$

During the mediation process, the evolution of the offers from supplier, in accordance to these optimal result, can be described by :

$$g_{1,1} = N_{max,1} \exp(-a\,e^{-bt}) \tag{14}$$

$$g_{1,2} = \text{Re}\{(S(t) - C_{alt}N_{max,1})/C_{alt}\} \exp(-a\,e^{-bt})$$

$$g_{2,1} = \text{Re}\{R(t) - [(C_{alt} - C_{init}) \times (N_{max,1} + (S(t) - C_{alt}N_{max,1})/C_{alt})]\} \exp(-a\,e^{-bt})$$

And therefore, the value of the offers received for the buyer is :

$$G(g_{n,m}) = \frac{V_1 g_{1,1}}{1+ru_1} + \frac{V_1 g_{1,2}}{1+ru_2} + \frac{V_2 g_{2,1}}{1+ru_1} \tag{15}$$

In the above example, the value of $N_{min}(t)$ is expected to decrease as a function of time, when the buyer's introspection leads to a reassessment of its requirements. We can expect a settlement to be reached when the supplier's offers meet the buyer's minimal requirements, therefore when $(g_{1,1} + g_{1,2}) = N_{min}(t)$.

In the case of the reverse condition :

$$\frac{V_1/V_{init}}{1+ru_2} < \frac{V_2}{1+ru_1} \tag{16}$$

Then optimisation of the $G(z_{n,m})$ function will instead favour the maximum reimbursement, as long as the minimum number of units $N_{min}$ is met. In this case, the results of linear optimisation would be:

$$g_{1,1} = N_{max,1} \exp(-a\, e^{-bt}) \qquad (17)$$

$$g_{1,2} = \text{Re}\{ N_{min}(t) - N_{max,1} \} \exp(-a\, e^{-bt})$$

$$g_{2,1} = \text{Re}\{ R(t) - [(C_{alt} - C_{init}) \times N_{min}(t)] \} \exp(-a\, e^{-bt})$$

Settlement is expected to occur for $(g_{1,1}+g_{1,2})=N_{core}$. Optimisation should therefore only occur for $g_{2,1}(t)$. The optimal solution will be the maximum amount of $g_{2,1}(t)$ that can be reached after the $N_{core}$ is met by the offers of the supplier, which is expected to occur when the $N_{mod}$ will have completely phased-out at $kt=\pi/2$. The value of the offers received for the buyer $G(g_{n,m})$ can be calculated once again using the function $G(g_{n,m})$ in (15).

## 4. RESULTS

Using the same illustrative case as in section 3.5 above and where the (12) condition is fulfilled, the results of $g_{1,2}+g_{1,2}$ and $N_{min}(t)$ can be represented over the mediation period by the following figures 2, 3, and 4.

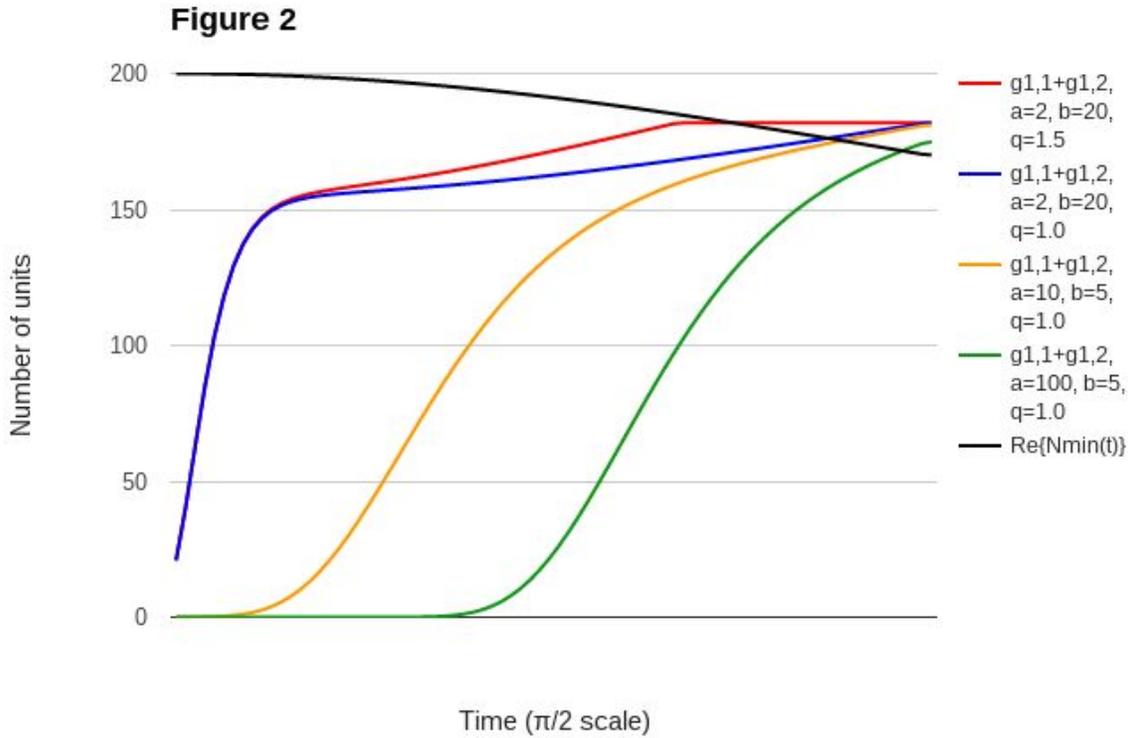

Figure 2 shows various examples with $V_1 = V_{init} = \$1000$, $V_2 = 0.8$, $C_{init} = \$700$, $C_{alt} = \$1100$, $N_{max,1} = 100$, $N_{core} = 170$, $N_{mod} = 30$, $S_{core} = \$200,000$, $S_{mod} = -\$30,000$, $k=1.0$. Since $N_{max,1} < N_{core}$, the minimal requirement of the buyer cannot be met in a single instalment of units deliveries. Therefore, the offers of each parties will meet to when the $\text{Re}\{N_{min}(t)\}$ line cross the $(g_{1,1}+g_{1,2})$ line, leading probably to a settlement corresponding to this value. It is to be noted that this occurs before $t$ reaches complete phase-out of the modulation at $\pi/2$.

For the red curve, $q=1,5$, $a=2$ and $b=20$, the parties positions meet at approximately $t \cong 1.16$, for 182 units. This is also the optimal solution for core constraints without mediation time-dependency (no Gompertz function applied to the offers of the supplier and considering only the core constraints of each party, but taking into consideration the discount $D(u_m)$ for all alternative performances instalments). When a settlement is reached at $t \cong 1.16$, the value of the reduction offered $g_{2,1}$ is \$27,272, approximately the same result as without any mediation time-dependency.

For all other parameters sets in figure 1, the number of units of the outcome is lower, therefore sub-optimal compared to the solution without mediation time-dependence. Early assessment of constraints *and* early concessions by the defaulting supplier yields the highest $F(z_i)$. These conclusions were found to be applicable with different sets of parameters where the condition (12) is met, and also for situations whereas $\text{Re}\{S(t)\} > \text{Re}\{C_{init}N_1 + C_{alt}(N_{min}(t) - N_{max,1})\}$ and where condition (16) is met.

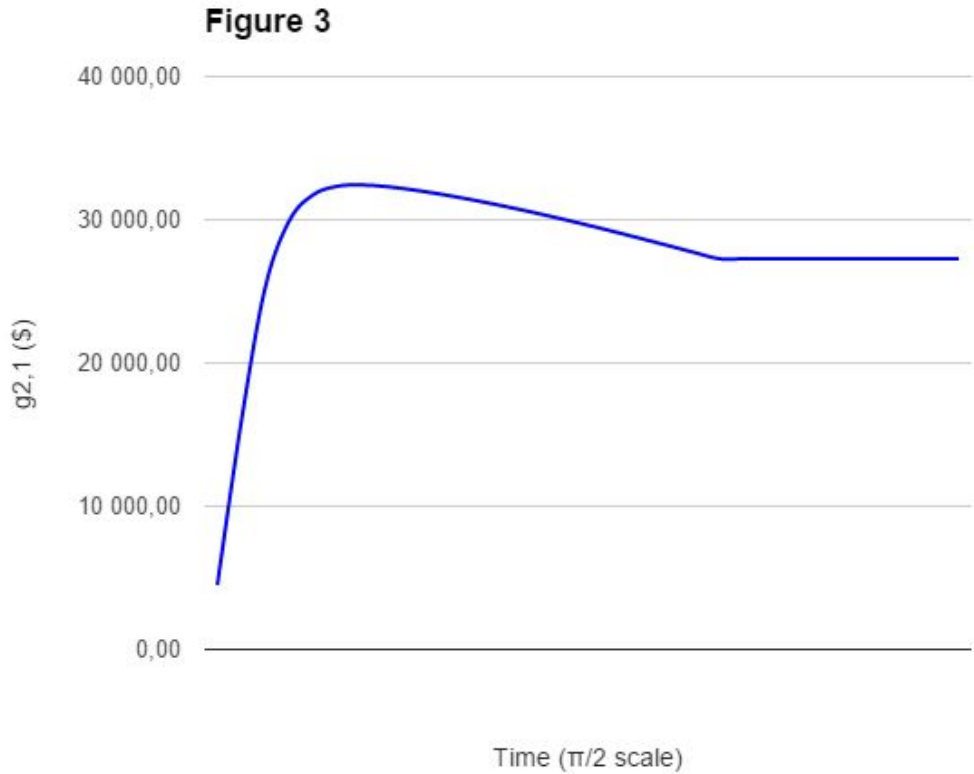

Figure 3 shows $g_{2,1}(t)$ using the same parameters as the best solution of figure 2 ($a=2$, $b=20$, $q=1.5$), with $R_{core}=\$100,000$ and $R_{mod}=-\$5,000$. It shows the value of the reimbursement offered by the supplier to settle the situation. At $t \approx 0.30$, the amount offered $g_{2,1}(t)$ peaks at \$32,432. Even by applying a Gompertz function to offers, the supplier may want to rapidly offer a larger amount of money to settle the issue, as it is less costly for this supplier than buying more expensive alternative units which cost 10% more than the contractual value V of the units. However, when a settlement is reached for the number of units at $t \approx 1.16$, the value of the reimbursement offered $g_{2,1}$ will have decreased to \$27,272.

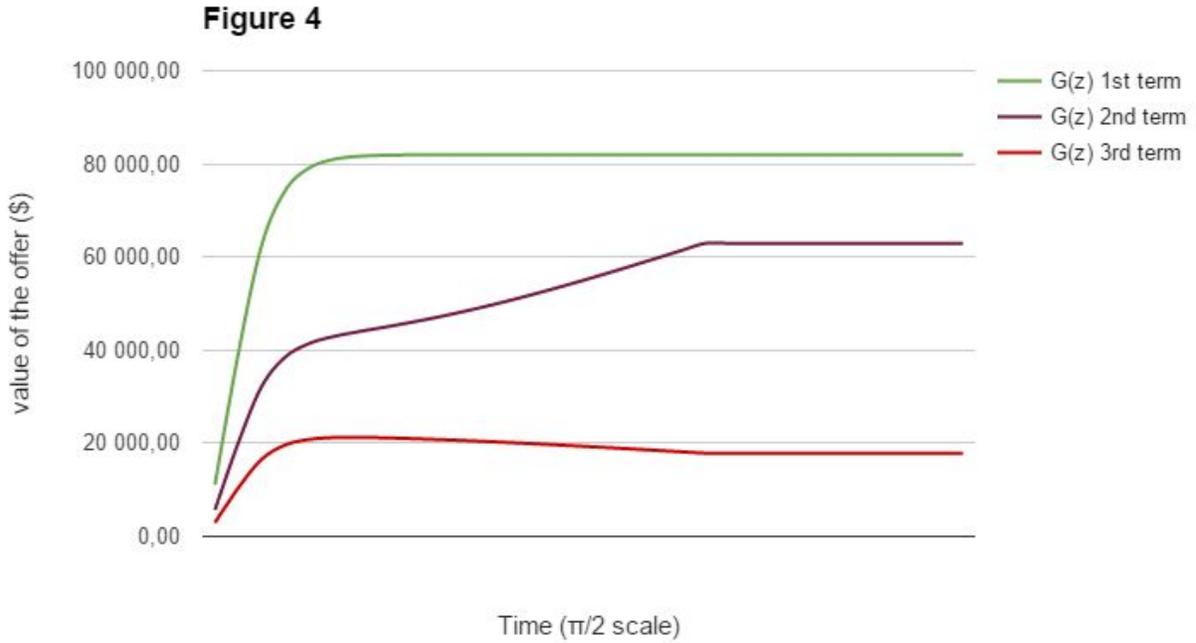

Figure 4 represent the calculated value of the three terms of the G($g_{n,m}$) in (15) as a function of the offers over the duration of the mediation process, for the set of parameters used in figure 2, and $r=0.2$, $u_1=1.1$ and $u_2=1.5$. The 1st term is the buyer's perceived value of the offer for the first instalment of alternative units $z_{1,1}$. At the moment of settlement $t \cong 1.16$, this 1st term is approximately $81,967, this is clearly the most valuable term. The 2nd term represent the perceived value of the offers for the second term, corresponding to the second instalment of alternative units $z_{1,2}$. At the moment of settlement, this 2nd term is approximately $62,937. The third term is the value of the reimbursement $z_{2,1}$. At the moment of settlement, this 3rd term is approximately $17,884. The total **max** G($g_{n,m}$) is therefore approximately $162,788. For comparison, the value $V_{init} \times N_{core} = \$170,000$, and the value $V_{init} \times (N_{core}+N_{mod}) = \$200,000$, both of which are higher than **max** G($g_{n,m}$). This is the expected result, the proposed model aiming at solutions as close as possible, within the constraints, to the initial value of the contract $V_{init} \times (N_{core}+N_{mod})$.

In the above examples, the discount rate due to delay in delivery of the second instalment was set in accordance with condition (12) to favour the delivery of a maximum number of alternative units in the second instalment over a potential reimbursement. If, using the same set of parameters changing only $u_2=4$, then it is condition (16) which is met. The optimal result would therefore be $g_{1,1}=100$, $g_{1,2}=70$, and $g_{2,1}=\$32,000$. This yields to **max** G($g_{n,m}$) = $148,405. This lower value compared to the situation of figure 4 issues from the higher discount due to the long delivery delay $u_2$.

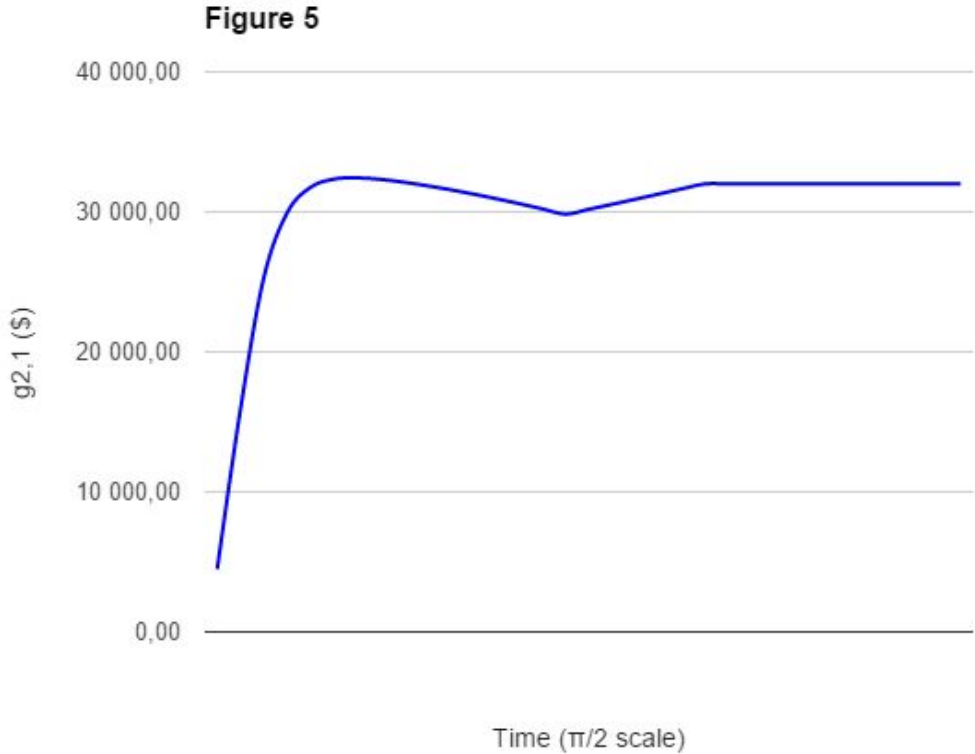

Figure 5 show the evolution of the $g_{2,1}$ during the mediation with $u_2=4$. The curve shows an initial rapid increase, with a peak at $t\cong 0.30$, then a progressive decrease a the offer of alternative units $g_{1,2}$ increases, decreasing the amount of money $R(t)$ available. Then, since the optimality favours reimbursement $g_{2,1}$ over alternative units $g_{1,2}$, the former increases again as the defaulting supplier's $R_{mod}$ and $S_{mod}$ phase-out, and stabilises at \$32,000 once $qt \geq \pi/2$. In order to achieve an optimal result, which corresponds to the complete phase-out of $R_{mod}$ and $S_{mod}$, it is however necessary to have $q>1$. Trivially, the inverse would lead to suboptimal $g_{2,1}$ results at the end of the mediation session $t=\pi/2$. Therefore, early assessment of its constraints by the defaulting supplier is again a requirement to achieve optimal results.

## 5. DISCUSSION

In this defaulting supplier problem, an optimal settlement requires an introspective analysis by each party of its own constraints (phase-out of the modulation to the core constraints). Without introspection and reassessment by each party, a suboptimal outcome, at best, may be achieved, which a party might disavow afterward despite entering into a settlement agreement. In some other cases, the partial presence of a modulation on top of the core constraint would even be sufficient to prevent any solution. The role of the mediator is therefore to facilitate this introspection by each party in order for the solution to correspond to a real optimal solution, not a hastily perceived solution.

However, as shown in the results above, the optimal settlement is achieved in situations where the defaulting supplier makes rapid concessions *and* assesses its own constraints earlier than the non-defaulting buyer.

In situations where the supplier is near insolvent, more specifically unable to perform its obligations to the buyer, and also unable to provide sufficient alternative performances sufficient to fulfil the contractual value of the initial contract, the function $F(z_{n,m})$ defined herein will be lower than this initial contractual value. Its maximisation will correspond to the highest value that the defaulting supplier is able to achieve. Also taking into account the negotiation process within mediation, we have defined a function $G(z_{n,m})$ which can be used to measure "optimality" of the offered solution and compare it to other possible outcomes. The offer which is closest to the initial contractual agreement will achieve highest $G(z_{n,m})$. This maximised $G(z_{n,m})$ will correspond to an acceptable solution for the buyer, and ideally, a settlement will issue with the corresponding $g_{n,m}$ values.

Where replacement suppliers are available, then the $G(z_{n,m})$ value could be used to take decisions using the comparative criteria of inequality (9). However, the buyer should not expect to be able to draw that comparison from the initial mediation position of the defaulting supplier, but should wait for the offers from the supplier to tend to their maxima after some time spent in mediation.

## **6. CONCLUSION**

The approach presented herein may be used as a tool for mediators to analyse typical mediation for cases where a supplier is in default to limitation in its solvency.

The mediator may gather constraints on each side, then propose an optimal solution or guide the parties to it, which solution will achieve the highest possible adherence to the initial contractual value. This, in turn, promotes contractual stability, while allowing the survival of the defaulting supplier. For the buyer, this is typically the best possible outcome in situations where the supplier, even under forced execution by the courts, will not be able to provide the initial value to the contract. Seeking a solution which corresponds the maximum performance of the defaulting supplier allows the buyer to obtain instead the maximum value, while also meeting its minimal operational requirements (constraints).

The time-dependant model presented herein shows that a mediator, in order to attain such optimal solution in a settlement agreement, should rapidly entice the defaulting party to seek alternatives to the contractual elements at issue and to reassess its internal constraints. This should be prioritised over the reassessment of the constraints of the non-defaulting party.

This approach could eventually be generalised to more complex contractual relationships. For example, a similar approach could be considered in situation whereas both parties are unable to fulfil their obligation under the contractual agreement.

*(Please send comments to the author by e-mail)*